\documentstyle[11pt,paspconf,epsfig]{article}

\begin{document}

\title{Blandford-Znajek process as a gamma ray burst central engine}

\author{Hyun Kyu Lee\altaffilmark{1}, R.A.M.J. Wijers, and G.E. Brown}
\affil{Department of Physics and Astronomy, State University of New York, 
Stony Brook, NY 11794, USA}

\altaffiltext{1}{Department of Physics, Hanyang University, Seoul 133-791,
Korea}

\begin{abstract}
We investigate the possibility that gamma-ray bursts are powered by a 
central engine consisting of a black hole with an external magnetic field 
supported by  a surrounding disk or torus. The rotational energy of the 
black hole can be extracted electromagnetically as a Poynting flux, a mechanism 
proposed by Blandford and Znajek(1977). Recently observed magnetars   
indicate that some compact objects have very high magnetic fields,
up to $10^{15}$\,G, which is 
required to extract the energy within the duration of a GRB, i.e.,
in $1000$\,s or less. We demonstrate also that 
the Poynting flux need not be substantially dominated by the disk. 
\end{abstract}

\keywords{gamma-ray bursts, black holes, Blandford-Znajek mechanism, 
Poynting flux, 
accretion disk}

\section{Introduction}
 
Gamma ray bursts presently provide great excitement in astronomy and
astrophysics as optical observations by way of many instruments give
considerable detail of the history of each burst. We are concerned
here with the prodigious energy in each burst, the estimate for GRB 971214
being  greater than $ 3 \times 10^{53}$ ergs (Kulkarni et al., 1998),
although this could be diminished if considerable beaming is involved in
the central engine, as we will discuss.
 
Amazingly, $2 \times 10^{54}$ ergs is just the rest mass energy of our sun,
so it seems immediately clear that the central engine for the GRB must be
 able to extract a substantial fraction of the rest mass energy of a compact
 object, neutron star or black hole, and convert it into energy of GRB.
 
The second criterion for the central engine is that it must be able to
deliver power over a long interval up to $\sim 1000$ seconds, since some
GRB's last that long, although other GRB's last only a fraction of a second.
It must also be able to account for the vast diversity in pulses, etc., or,
alternatively, one must have a number of diverse mechanisms.

We believe the need to deliver power over the long time found in some bursters
to be the most difficult requirement to fulfill, since the final merger
time of the compact objects is only a fraction of a second and it is difficult
 to produce a high energy source of, e.g., $\nu \bar{\nu}$-collisions that
goes on for more than two or three seconds.
 
For many years mergers of binary neutron stars  and of a neutron star with
a black hole were considered to be likely
sources for the GRB's. The estimated merger rate in our Galaxy of a few GEM
(Galactic Event per Megayear) is of the right order for
the occurrence of GRBs. 
The possible problem with  the binary   mergers might be  
the ejected materials during the merging processes.  
Not more than $\sim 10^{-5}
M_{\odot}$ of baryons can be involved in the GRB, since it would not be
possible to accelerate a higher mass of nucleons up to the Lorentz factors
needed with the energies available. 

We find that the emergence  or presence of a black hole  during the 
merging process
to be  particularly attractive. 
The baryon number ``pollution" problem can be solved
by the main part of the baryons going over the event horizon. In the
Blandford-Znajek mechanism(Blandford \& Znajek 1977) 
we wish to invoke, a substantial proportion of
the rotational energy of the black hole, which will be sent into rapid
rotation by swallowing up the neutron star matter, can be extracted through
the Poynting vector(Meszaros \& Rees 1997). 
The rate of extraction is proportional to the square of
magnetic field strength, $B^2$, as we shall discuss, so that power can be
furnished over varying times, depending upon the value of $B$. With
substantial  beaming, we estimate that $B\sim 10^{15}$G would be sufficient
to power the most energetic GRBs with $\sim 10^{53}ergs$.

Failed supernovae were suggested by Woosley( Woosley 1993)
as a source of GRB. In this case the black hole would be formed in the
center of a massive star, and surrounding baryonic matter would accrete into
it, spinning it up. This mechanism is often discussed under the title of
hypernovae(Paczynski 1998).
 
More recently Bethe and Brown(Bethe \& Brown 1998) found that in binary
neutron star evolutions, an order of magnitude more low-mass black-hole,
neutron-star binaries were formed than binary neutron stars. The low-mass
black-hole mass of $\sim 2.4M_{\odot}$ looks favorable for the
Blandford-Znajek mechanism.
 
In some calculations which begin with a neutron star binary, one of the
neutron stars evolves into a black hole in the process of accretion, and the
resulting binary might also be a good candidate for GRBs. In any case, there
 are various possibilities furnished by black-hole, neutron-star binaries.

The structure we are proposing as a central engine of the GRB is a system of
a black hole
surrounded by centrifugally supported material, either an accretion disk or
a debirs torus left from a recently disrupted object.
The rotating black hole is threaded by a strong
magnetic field. Along the baryon-free funnel relativistic jets fueled by
Poynting outflow give rise to the GRB.  The interaction between disk and black
hole is characterized by accretion  and magnetic coupling.
We consider  the BZ process only
after the main accretion process is completed, leaving an accretion disk of
cold  residual material, which can support a strong enough magnetic field.   

\section{Blandford and Znajek process}

 Two decades ago Blandford and Znajek(Blandford \& Znajek 1977) proposed a
process(BZ) in which rotational energy of a black hole can be efficiently
extracted.    
Consider a half hemisphere(radius $R$) rotating with angular velocity $\Omega$
and a  circle on the surface at fixed $\theta$( in the spherical polar
 coordinate
system) across which a surface current $I$ flows down from the pole. When the
external magnetic field $B$ is imposed to thread the surface outward normally,
the surface current  feels a force and  the
torque due to the Lorentz
force exerted by the annular ring of  width $Rd\theta$ is
\begin{eqnarray}
d T 
  = - \frac{I}{2\pi} d \Psi. \label{torque}
\end{eqnarray}
where $d \Psi$ is the magnetic flux through annular ring extended by
$d\theta$. 
 From this magnetic braking, we can calculate the rotational energy loss rate
\begin{eqnarray}
P_{rot} = \frac{1}{2\pi} \int \Omega I d \Psi. \label{dprot}
\end{eqnarray}
 
Blandford and Znajek demonstrated that such a  magnetic braking
is possible, provided that the external charge  and current distributions 
around the black hole can support the force-free condition. 
The Blandford-Znajek process has been reformulated in the frame work of the
membrane paradigm(Thorne, Price \&MacDonald 1986)
 in which the complicated physics near the horizon can be expressed in terms
physical quantities defined on the stretched horizon. For the axial symmetric 
and force-free magnetosphere the rotational energy loss rate, $P_{rot}$, and
the out going power, $P_{BZ}$, are given respectively by
\begin{equation}
P_{rot} = \frac{1}{2\pi} \int \Omega_H I d \Psi, \,\, 
P_{BZ} =  \frac{1}{2\pi}\int \Omega_{F}
 I d \Psi
\end{equation} 
for a black hole with mass $M$ and angular momentum $J$(angular velocity
$\Omega_H = \frac{J}{2M^2 r_H}$) with rigidly rotating magnetic fields
(angular velocity $\Omega_F$). 

The maximum amount of energy which can be extracted out of the
black hole
without violating the second  law of thermodynamics  is the rotational
energy, $E_{rot}$, which is defined as
\begin{equation}
E_{rot} = Mc^2 - M_{irr}c^2 = (1- \sqrt{[1+\sqrt{1-\tilde{a}^2}/2]})Mc^2 
\end{equation}
where
$M_{irr} = \sqrt{\frac{A_H c^4}{16 \pi G^2}}$
and
$A_H$ is the surface area of the black hole(equivalently the black 
hole entropy,  
$S_H = \frac{k_B A_H}{4\hbar}$) 
and $ \tilde{a}= \frac{Jc}{M^2G}$ is the rotation parameter.

For a maximally rotating black hole ($\tilde{a}=1$),
$E_{rot} =0.29Mc^2$.
However in the physical process a fraction  of the rotational energy is 
dissipated into the black
hole
increasing the
entropy or equivalently irreducible mass:$P_{diss} =
P_{rot} - P_{BZ} = \frac{1}{2\pi} \int (\Omega_H - \Omega_F) I d\Psi$.
For the optimal process,
$\Omega_F \sim 0.5 \Omega_H$,  
the total Blandford-Znajek  power is given by(Lee, Wijers \& Brown 1999)  
\begin{eqnarray}
P  = 1.7\times 10^{50}  \tilde{a}^2 f(\tilde{a})(\frac{M}{M_{\odot}})^2
    (\frac{<B_H>}{10^{15}gauss})^2  \, erg/s, \label{powerm}
\end{eqnarray}
where  $f(\tilde{a}) = 0(\tilde{a} =0) \rightarrow 1.14(\tilde{a} =1)$
and the total energy extracted out is calculated(Phinney 1983, Okamoto 1992)
 to be
\begin{equation}
E_{BZ} = 0.091 \, Mc^2 \, = 1.6 \times 10^{53}(\frac{M}{M_{\sun}}) erg,
\end{equation}  
and the time scale $\tau_{BZ} \sim 10^3(\frac{10^{15} gauss}{B_H})^2 
(\frac{M}{M_{\sun}})$s.
 
The fluences of the recently observed GRB971214 (Kulkarni et al.
 1998)
and GRB990123 (Kulkarni et al. 1999) 
correspond to
$E_{\gamma} =
[10^{53.5},  3.4 \times 10^{54}]
(\frac{\Omega_{\gamma}}{4\pi}) erg$ which 
are consistent with $E_{BZ}$ if the considerable  beaming is considered.
Also the durations of gamma ray bursts  suggest
that if a strong enough magnetic field ($\sim 10^{15}$G)
on the black  
hole can be supported by the surrounding material(accretion disk) the BZ
process is  a good candidate to provide the powerful energy of the
GRB in the
observed time interval up to 1000s , which is comparable to the BZ time scale
$\tau_{BZ}$.
 
The rotation parameter of the black hole can be estimated
by assuming a substantial part of the angular momentum
of the
merging compact binary systems
or collapsars( a fraction($x$) of the specific angular
momentum)
 can be imparted onto the black hole when a fraction 
($y$) of the mass
collapses into the black hole.
 From the semiquantitative estimation(Lee, Wijers \& Brown 1999)
 we get
$\tilde{a} = 0.53\frac{x}{y}$ for BH-NS merger with
$M_{BH}=2.5M_{\odot}, M_{NS}= 1.5M_{\odot}$  and the  tidal radius
$\sim 10^6$cm,
$\tilde{a}
     = 0.67\frac{x}{y}$ for NS-NS merger, and $\tilde{a} =
2.3\frac{x}{y}$ for collapsars with
 massive rapidly spinning
progenitors
($M_o \sim 40M_{\odot}$).
Hence it is very plausible to have a rapidly rotating black hole
as a resulting  object in the center in the merging
systems and also in hypernovae of large angular momentum progenitors, but
a precise value of $\tilde{a}$ will be difficult to calculate.

\section{Magnetized accretion disk}
 The
presence of the accretion disk is important for the BZ process  because it is
the
supporting system of the strong
magnetic field on the black hole, which would disperse without the pressure
from the fields anchored in the accretion disk.
Recent numerical calculations(Popham,  Woosley \& Fryer 1999) show
that accretion disks formed by various  merging processes are found to have
large  enough
pressure such that they can support $\sim 10^{15}G$ assuming a value of the
disk viscosity parameter $\alpha
\sim 0.1$, where $\alpha$ is the usual parameter in scaling the viscosity. 

The discovery that soft gamma ray
bursts are magnetars(Kouveliotou et al. 1998)
also supports the presence of the strong magnetic fields of
$\sim 10^{15}G$ in nature. It is also possible that existing magnetic fields
can be increased by the dynamo effect.   
The identification by now of
three soft gamma repeaters as strong-field pulsars indicates that there may be
a large population of such objects:
since the pulsar spindown times scale as $B^{-1}$,
we would expect to observe only 1 magnetar for every 1000 normal pulsars if they
were formed at the same rate, and if selection effects were the same for the two
populations. We see 3 magnetars and about 700 normal pulsars, but since they
are found in very different ways the selection effects are hard to quantify.
It is nonetheless clear that magnetars may be formed in our Galaxy at a rate
not very different from that of normal pulsars.  

The life time of the accretion disk determined by the accretion rate 
is also very important
for the GRB time
scale because it supports the magnetic field on the black hole. 
The strong magnetic field perpendicular to the disk produces a high
 accretion rate $\dot{M} =
\sim 10^{-4} M_{\sun}s^{-1}$(Lee, Wijers \& Brown 1999).
However  the total accretion is less than 
$10^{-1}M_{\sun}$ during the GRB time scale and it may not change the
disk structure for the Blandford-Znajek process significantly.    
According to
numerical simulations of
merging systems which evolve eventually into black hole - accretion disk
configuration (Popham,  Woosley \& Fryer 1999) the viscous life times 
are 0.1 - 150 s, which are
not inconsistent with the GRB time scale.  
Also it has been pointed out (Meszaros, Rees \& Wijers 1998)
that a residual cold disk of $\sim 10^{-3} M_{\odot}$ can support $10^{15}G$, 
even
after the major part of the accretion disk has been drained into the black hole
 or dispersed away.
 
The current conservation condition, namely that the total current flows onto the black hole 
should go into the inner edge($r_{in}$) of the accretion disk, implies 
\begin{equation}
2M B_{\phi}^H(\theta=2\pi) = \tilde{\omega}(r_{in})B^{disk}_{\phi}(r_{in}).
\end{equation}
 Since the cylindrical radius in Kerr geometry $\tilde{\omega}(r_{in}) 
> 2M $, we can see that $B_{\phi}^H$ is larger than 
$ B^{disk}_{\phi}$.
 From the boundary conditions on the horizon in the optimal case 
and on the accretion disk (Blandford 1976) with angular
velocity $\Omega_D$ respectively
\begin{equation}
B_{\phi}^H = -\Omega_H M B_H, \, \, B_{\phi}^{disk} = 
-2 \Omega_D r B_z^{disk},
\end{equation}
we get 
\begin{equation}
\frac{B_z(r_{in})}{B_H} =  \sqrt{\frac{GM}{r_{in}c^2}}
\frac{\tilde{a}}{2}\frac{G M}{r_Hc^2} < 1,   
\end{equation}
and the power of the disk  magnetic braking $P_{disk}
= \frac{2}{f(\tilde{a})}(\frac{GM}{r_Hc^2})^2 P_{BZ}$. 
It shows that the magnetic field on the horizon cannot be smaller than 
that on the inner edge of the accretion disk and the disk power need not
be substantially larger than that from the black hole (Livio, Ogilvie 
\& Pringle 1998; 
Ghosh \& Abramowicz 1997; Lee, Wijers \& Brown 1999 ). 

The energy outflow from the disk is mostly directed vertically from 
the disk where the baryon loading is
supposed to be relatively high enough to keep the baryons
 from being highly relativistic.
Therefore the BZ process from  the disk
can be considered to have  not much to do with gamma ray
burst phenomena. However the BZ from the disk could power an outflow with lower
$\Gamma$, but nonetheless high energy, which could cause an afterglow
at large angles. That would lead to more afterglows being visible than GRBs,
because the afterglows are less beamed.

\section{Conclusion}
We have evaluated the power and energy that can be extracted from
a rotating black hole immersed in a magnetic field, the 
Blandford-Znajek process. We improve on earlier calculations to find that
the power from a black hole of given mass immersed in an external field is
ten times greater than previously thought. The amount of energy that can
be extracted from a black hole in this way is limited by the fact that
only 29\% of the rest mass of a black hole can be in rotational energy,
and that the optimal efficiency with which energy can be extracted from
the hole via the Blandford-Znajek effect is 31\%. The net amount of energy
that can be optimally extracted is therefore 9\% of 
the rest energy of the black
hole. We consider various scenarios for the formation of rotating black
holes in gamma-ray burst engines, and while the resulting angular momenta
are quite uncertain in some cases, it seems that the required values of
the rotation parameter, $\tilde{a} > 0.5$, are achievable.

The rate at which angular momentum is extracted depends on the magnetic
field applied to the hole. A field of $10^{15}$\,G will extract the
energy in less than 1000\,s, so time scales typical of gamma-ray bursts
can be obtained.  Since the black hole cannot carry a field, there must
be an ambient gas in which the field is anchored that drives the Poynting
flux from the black hole. The most obvious place for it is the accretion
disk or debris torus surrounding the black hole just after it formed.
 
It has been argued  that a field turbulently generated
in the disk would not give a strong Blandford-Znajek flux (Livio, Ogilvie
 \& Pringle 1998), because
the disk would dominate the total Poynting output, and no more than the
disk's binding energy could be extracted. We show explicitly that a field
in the disk could be much greater, for example if it is derived from the
large, ordered field of a neutron star that was disrupted. The
field distribution proposed by Blandford (Blandford 1976) for such a case
would allow
the field on the hole to be larger  than on the disk, such that the
Poynting flow would not be dominated by the disk  and not subject to any
obvious limits imposed by the disk.

Recent hydrodynamic simulations (Ruffert \& Janka 1998)
of merging binaries
show that along the rotation axis of the black hole an almost
baryon-free funnel is possible. This can be easily understood since the
material
above the hole axis has not much angular momentum so that it can be drained
quickly, leaving a baryon-free funnel.
Hence relativistically expanding
jets along the funnel, fueled by Poynting outflow which is  collimated along the
 rotation axis,  can give rise to gamma ray
bursts effectively.  Recent observation (Kulkarni et al. 1999) seems to
provide an evidence of beaming for GRB990123. It is also interesting to
note that the beaming via Blandford-Znajek process with the possible 
disk precessing  can be applied to
simulate the temporal structures of the gamma ray bursts
(Portegies-Zwart, Lee \& Lee 1999).

We also note that a Poynting flow may provide an alternative way of providing
a very large magnetic field for the shocked material that radiates the
afterglow and the gamma-ray burst itself: the standard assumption is that
the required high fields grow turbulently in the shocked gas, up to
near-equipartition values. But the field in the Poynting flow only decreases
as $1/r$, so if it is $10^{15}$\,G at $r=10^5$\,cm, it could be as high as
$10^{4}$\,G at the deceleration radius ($10^{16}$\,cm), ample to cause an
energetic gamma-ray burst.

We are grateful to Roger Blandford, Sterl Phinney and Kip Thorne for
guidance and useful discussions. This work is supported by  
the U.S. Department 
of Energy under Grant No.  DE-FG02-88ER40388. HKL is supported 
also in part by KOSEF-985-0200-001-2 and  BSRI-98-2441.


\begin{references}


\reference Bethe, H \&  G.E. Brown, G.E. 1998, ApJ 506, 780
\reference Blandford, R.D. 1976, MNRAS 176, 465
\reference Blandford, R.D., \& Znajek, R.L. 1977, MNRAS 179, 433
\reference Ghosh, P. \& Abramowicz. M.A. 1997, MNRAS  292, 887
\reference Kouveliotou, C. et al. 1998, Nature  393, 235
\reference Kulkarni, S.R. et al. 1998, Nature, 393, 35
\reference Kulkarni, S.R. et al. 1999, astro-ph/9902272 v2
\reference Lee, H.K., Wijers, R.A.M.J. \& Brown, G.E. 1999, 
in preparation 
\reference Livio, M., Ogilvie, G.I. \& Pringle, J.E. 1999,
ApJ 512, 100L 
\reference Meszaros, P., Rees, M.J.1997, ApJ 482, L29 
\reference Meszaros, P., Rees, M.J., \& Wijers, R.A.M.J. 1998
    astro-ph/9808106
\reference Okamoto, I. 1992, MNRAS 294, 192
\reference Paczynski, B. 1998, ApJ 494, L45
\reference Phinney, S. 1983, PhD thesis
\reference Popham, R., Woosley, S.E., \& Fryer, C.L.
 1999, astro-ph/9807028  
\reference Portegies-Zwart,S., Lee, C.-H. \& Lee, H.K. 1999,
ApJ in press(astro-ph/9808191)
\reference Ruffert,M. \&  Janka,H.-Th. 1998, astro-ph/9804132
\reference Thorne, K.S., Price, R.H. \&  MacDonald, D.A. 1986,
Black Holes;
The Membrane Paradigm 
\reference Woosley, S.E. 1993, ApJ  405, 273 

\end{references}
\end{document}